\documentstyle[epsf,12pt]{article}

\begin{document}

\begin{titlepage}

\parindent=12pt
\baselineskip=20pt
\textwidth 15 truecm
\vsize=23 truecm
\hoffset=0.7 truecm

\noindent{May 1995}
\hfill{LPTHE Orsay 95--39}\par
\vskip .5 truecm
\centerline{\bf FLOW EQUATIONS FOR THE RELEVANT PART}\par
\centerline{\bf OF THE PURE YANG--MILLS ACTION} \par
\vskip 1 truecm
\centerline{{\bf Ulrich ELLWANGER}\footnote{e--mail:
ellwange@qcd.th.u--psud.fr}}
\par
\centerline{Laboratoire de Physique Th\'eorique et Hautes
Energies\footnote{Laboratoire
associ\'e au Centre National de la Recherche Scientifique -- URA 63}} \par
\centerline{Universit\'e de Paris XI, b\^atiment 211, 91405 Orsay Cedex,
France} \par
\vskip .5 truecm
\centerline{{\bf Manfred HIRSCH}\footnote{Supported by a DAAD--fellowship HSP
II
financed by the German Federal Ministry for Research and Technology; e--mail:
hirsch@qcd.th.u--psud.fr} and {\bf Axel WEBER}\footnote{Supported by a
LGFG--fellowship of the Land Baden--W\"urttemberg
and a DAAD--fellowship; e--mail: weber@qcd.th.u--psud.fr}}\par
\centerline{Institut  f\"ur Theoretische Physik}\par
\centerline{Universit\"at Heidelberg, Philosophenweg 16, D--69120 Heidelberg,
FRG}\par
\centerline{and}\par
\centerline{Laboratoire de Physique Th\'eorique et Hautes
Energies\footnotemark[2]} \par
\centerline{Universit\'e de Paris XI, b\^atiment 211, 91405 Orsay Cedex,
France} \par

\vskip 2 truecm
\noindent \underbar{{\bf Abstract}} \par
Wilson's exact renormalization group equations are derived and
integrated for the relevant part of the pure Yang--Mills action. We discuss in
detail
how modified Slavnov--Taylor identities controle the breaking of BRST
invariance in the
presence of a finite infrared cutoff $k$ through relations among different
parameters in
the effective action. In particular they imply a nonvanishing gluon mass term
for
nonvanishing $k$. The requirement of consistency between the renormalization
group flow
and the modified Slavnov--Taylor identities
allows to controle the self--consistency of truncations of the effective
action.
\par

\end{titlepage}

\section{Introduction}

One of the most important and most interesting problems in quantum field theory
today
is the description of non--perturbative phenomena. The main tools used so far
to
tackle these problems are Schwinger--Dyson equations and the formulation on a
(finite)
space--time lattice. The difficulties with both methods can be traced back to
the
fact that in quantum field theory effects on all length scales are involved in
an essential way.

The last years have witnessed the emergence of a new technique which may be
capable
of surmounting these difficulties. It originated much earlier in the work of
Wilson et
al.\ \cite{Wilson}, and is based on the idea that the effects of different
length
scales should be taken into account one after the other rather than all at a
time.
The resulting `exact renormalization group equations' or `flow equations' are
formulated directly in continuous Euclidean space--time, thereby avoiding any
lattice
artifacts. Polchinski \cite{Pol} was the first to present a version adapted to
the
needs of quantum field theory. His exact renormalization group equation
determines the
flow of the effective Lagrangian;
by taking the Legendre transform a flow
equation for the effective action has been obtained later
\cite{quantact}--\cite{E2}.
The effective action being the generating functional of one--particle
irreducible Green
functions, this formulation is more suited to the applications we have in mind
than
the older one involving the effective Lagrangian, which generates the truncated
connected Green functions.

Technically, the flow equation for the effective action is obtained by
introducing
an infrared momentum cutoff $k$ which is then varied continuously. At a large
scale
$\bar{k}$ only the quantum fluctuations on scales above $\bar{k}$ are
integrated
out, so that the corresponding effective action $\Gamma_{\bar{k}}$ incorporates
the
microscopic physics in the sense of an effective field theory. In the limit $k
\to 0$
the effective action becomes the usual one, generating the physical
one--particle
irreducible Green functions. The flow equation now provides a differential
equation
interpolating between $k = \bar{k}$ and $k = 0$, so given $\Gamma_{\bar{k}}$
one can
in principle calculate the physical effective action by just integrating the
flow
equation. Of course in practice it will not be possible to perform the
integration
exactly, since the functional $\Gamma_k$ will have to be described in general
by
infinitely many $k$--dependent parameters. What renders this approach useful,
then,
is the existence
of approximations which take into account only a finite number of parameters
and
provide nevertheless a sensible description of the relevant physical effects.

Historically, flow equations have been used first to simplify proofs of
perturbative
renormalizability \cite{Pol},\cite{Bon1},\cite{pert}--\cite{BDM}. Later they
have been
employed for the
description of non--perturbative phenomena like phase transitions, critical
exponents, bound states and condensates \cite{E2,W1}. Of course it is very
desirable to apply this method to gauge theories like QCD. However one
encounters here
the obvious problem that the infrared cutoff $k$ breaks the gauge or BRST
invariance.
Recently two different ways to cope with this problem have been proposed. In
\cite{W2}
background fields have been introduced in order to maintain gauge invariance.
However the resulting additional dependence of the effective action $\Gamma_k$
on the background field leads
to new difficulties. In \cite{B}--\cite{E1} it has been suggested to use
modified
Ward or Slavnov--Taylor identities to treat the symmetry breaking induced by
the
infrared cutoff. These identities guarantee the gauge or BRST invariance of the
physical effective action at $k = 0$. We follow here the formulation presented
in
\cite{E1}.

The aim of the present paper is to demonstrate the suitability of this approach
to
calculations in gauge theories. For definiteness we consider SU(3)-Yang-Mills
theory,
but the concept is by no means limited to this particular case. The emphasis
here is
on the method, so we shall work with a rather simple but
nevertheless non--trivial approximation for the $k$--dependent effective
action. However,
the formalism is general and can be combined with other techniques already
developed in
connection with the flow equations.

The paper is organized as follows: In section 2 we present the flow equations
and
modified Slavnov--Taylor identities following \cite{E1}, but using a more
concise
and generalizable notation. In section 3 we outline how this framework can be
employed
to calculate the physical effective action, and we present the approximation we
shall
use in the following. Section 4 represents the main part of the paper and
contains
the results from the integration of the flow equations. We shall find in
particular
that the BRST invariance of the original theory provides us with the means to
control
the approximation we have introduced. Section 5 contains our conclusions and a
short
outlook.

\section{Flow equations and modified Slavnov--Taylor \newline identities}

Throughout this paper we consider
SU(3)--Yang--Mills theory in four--dimen\-si\-o\-nal Euclidean space--time,
although the extension to general SU(N)--Yang--Mills theories in
$d$--dimensional space--time does not involve additional complications.

We take the classical action to be
\begin{eqnarray} S & = & \int\mbox{d}^4
x\left\{{\textstyle\frac{1}{4}}F^a_{\mu\nu}
   F^a_{\mu\nu} + \frac{1}{2\alpha}\partial_\mu A^a_\mu\partial_\nu A^a_\nu
   + \partial_\mu\bar{c}^a\left(\partial_\mu c^a + g f^a{}_{bc}A^b_\mu c^c
   \right)\right.\nonumber \\ & & \left.
   {} - K^a_\mu\left(\partial_\mu c^a + g f^a{}_{bc}A^b_\mu c^c\right) - L^a
   {\textstyle\frac{1}{2}}g f^a{}_{bc}c^b c^c +
\bar{L}^a\frac{1}{\alpha}\partial_\mu
   A^a_\mu \right\} \label{class}
\end{eqnarray}
with
\begin{equation}
  F^a_{\mu\nu} = \partial_\mu A^a_\nu - \partial_\nu A^a_\mu
   + g f^a{}_{bc}A^b_\mu A^c_\nu \:,
\end{equation}
where we have included the usual gauge fixing and ghost parts and coupled
external
sources to the BRST variations
\begin{eqnarray} \delta A^a_\mu & = & \left(\partial_\mu c^a + g f^a{}_{bc}
   A^b_\mu c^c\right)\zeta \nonumber\\
   \delta c^a & = & {\textstyle\frac{1}{2}}g f^a{}_{bc}c^b c^c\zeta \nonumber\\
   \delta \bar{c}^a & = & -\frac{1}{\alpha}\partial_\mu A^a_\mu\zeta \:.
\label{BRST}
\end{eqnarray}
Here $\zeta$ is a Grassmann parameter. The invariance of $S$ at $\bar{L} = 0$
under BRST transformations can then be expressed as
\begin{eqnarray} 0 & = & \left.\int\mbox{d}^4 x\left\{\delta
A^a_\mu\frac{\delta S}
   {\delta A^a_\mu} + \delta c^a\frac{\delta S}{\delta c^a} + \delta\bar{c}^a
   \frac{\delta S}{\delta\bar{c}^a}\right\}\right|_{\bar{L} = 0}\nonumber \\
   & = & \zeta\left.\int\mbox{d}^4 x\left\{\frac{\delta S}{\delta K^a_\mu}
   \frac{\delta S}{\delta A^a_\mu} - \frac{\delta S}{\delta L^a}
   \frac{\delta S}{\delta c^a} - \frac{\delta S}{\delta\bar{L}^a}
   \frac{\delta S}{\delta\bar{c}^a}\right\}\right|_{\bar{L} = 0} \:.
\end{eqnarray}

To derive the flow equations we add an infrared momentum cutoff term
\begin{equation}
  \Delta S_k = {\textstyle\frac{1}{2}}A\cdot R_k\cdot A
   + \bar{c}\cdot\tilde{R}_k\cdot c
\end{equation}
to the action $S$. Here we have introduced a matrix notation which will be
useful
in the following. It is most easily visualized in momentum  space, where the
matrix product implies integrations over momenta as well as summations over
group and Lorentz indices. As an example, $A\cdot R_k\cdot A$ actually stands
for
\begin{equation}
  \int\frac{\mbox{d}^4 p}{(2\pi)^4}\int\frac{\mbox{d}^4 q}{(2\pi)^4}
  A^a_\mu(-p) R^{ab}_{k,\mu\nu}(p,-q) A^b_\nu(q) \:.
\end{equation}
Imposing energy--momentum conservation and global SU(3)--invariance we have
\begin{eqnarray}
  R^{ab}_{k,\mu\nu}(p,-q) &=& R_{k,\mu\nu}(p) \delta^{ab} (2\pi)^4\delta(p - q)
   \nonumber \\
  \tilde{R}^{ab}_k(p,-q) &=& \tilde{R}_k(p) \delta^{ab} (2\pi)^4\delta(p - q)
   \label{cutoff} \:.
\end{eqnarray}
The functions $R_{k,\mu\nu}(p)$ and $\tilde{R}_k(p)$ introduce smooth infrared
cutoffs $k$ in the gluon and ghost field propagators and vanish identically at
$k = 0$,
\begin{equation}
  R_{k,\mu\nu}(p),\: \tilde{R}_k(p) \to 0 \quad\mbox{as}\quad k \to 0 \:.
  \label{cond1}
\end{equation}
Furthermore,
\begin{equation}
  R_{k,\mu\nu}(p),\: \tilde{R}_k(p) \to 0 \quad\mbox{`rapidly' as}\quad
  p^2 \to \infty \:, \label{cond2}
\end{equation}
so that $R_k$ and $\tilde{R}_k$ act simultaneously as ultraviolet cutoffs in
the integrals on the r.h.s.\ of (\ref{floweq}) and (\ref{STI}) below.
Explicit expressions for $R_k$ and $\tilde{R}_k$ will be given later.

The $k$--dependent generating functional of connected Green functions $W_k$ is
defined by
\begin{equation} e^{W_k[J,\chi,\bar{\chi},K,L,\bar{L}]} = \int{\cal D}_{reg}
   (A,c,\bar{c})e^{-\Delta S_k - S + J\cdot A + \bar{\chi}\cdot c + \bar{c}
   \cdot\chi} \label{genf} \:,
\end{equation}
where we have attached the index ``reg'' to the functional integration measure
to indicate that the functional integral has been regularized in the
ultraviolet.
This regularization is necessary to give a meaning to (\ref{genf}), and it
should
be independent of $k$ and preserve the symmetries of $S$. It is not quite
clear whether such a regularisation exists in a non--perturbative sense.
In the following we shall
nevertheless assume its existence in order to present a simple
derivation of the flow equations and the modified Slavnov--Taylor identities
(STI) and to illustrate the physical meaning of the quantities under
consideration.
Once we have obtained the flow equations and modified STI, however, we can use
them
without reference to (\ref{genf}) and thus do not rely on the existence of an
appropriate regularization any more, as has been emphasized in \cite{E1}. In a
more rigorous sense, the theory may directly be defined via the flow equations
and
modified STI \cite{B}.

So let us, for the moment, use (\ref{genf}) in order to define the functional
$W_k$.
Since $\Delta S_k$ vanishes for $k = 0$, we
see from (\ref{genf}) that $W_0$ is the generating functional of the physical
connected Green functions.

For general $k$ we pass over to the Legendre
transform with respect to the sources $J,\chi,\bar{\chi}$,
\begin{equation} \Gamma_k[A,c,\bar{c},K,L,\bar{L}] = A\cdot J + \bar{\chi}\cdot
c
   + \bar{c}\cdot\chi - W_k[J,\chi,\bar{\chi},K,L,\bar{L}] \:,
\end{equation}
\begin{equation} \mbox{where}\quad A = \frac{\delta W_k}{\delta J}\:,\quad
   c = \frac{\delta W_k}{\delta\bar{\chi}}\:,\quad \bar{c} = -\frac{\delta W_k}
   {\delta\chi}\:.
\end{equation}

Finally, $\hat{\Gamma}_k$ is defined by
\begin{equation} \Gamma_k = \hat{\Gamma}_k + \Delta S_k \:.
\end{equation}
At $k = 0$, $\hat{\Gamma}_0 = \Gamma_0$ becomes the physical effective action
or
the generating functional of one--particle irreducible Green functions.

Differentiating $\hat{\Gamma}_k$ with respect to $k$ leads to the flow
equations.
To cast them into a concise and generalizable form, we introduce the
superfields
\begin{equation} \Phi = (c,\bar{c},A)\quad\mbox{and}\quad\bar{\Phi} =
(-\bar{c},c,A)
\end{equation}
and (for later use) the external supersources
\begin{equation} Q = (L,\bar{L},K) \label{ssource} \:.
\end{equation}
The flow equations finally read \cite{quantact}--\cite{E2}
\begin{equation} \partial_k\hat{\Gamma}_k =
{\textstyle\frac{1}{2}}\mbox{STr}\left[
   \partial_k {\cal
R}_k\cdot\left(\frac{\delta^2\hat{\Gamma}_k}{\delta\bar{\Phi}
   \delta\Phi} + {\cal R}_k\right)^{-1}\right] \:. \label{floweq}
\end{equation}
Here the inverse is to be taken in the matrix sense including the momenta as
indices.
The elements of the matrix of second derivatives are in our notation given by
\begin{equation}
\left(\frac{\delta^2\hat{\Gamma}_k}{\delta\bar{\Phi}\delta\Phi}
   \right)_{ij}(p,q) =
(2\pi)^8\frac{\delta^2\hat{\Gamma}_k}{\delta\bar{\Phi}_i(-p)
   \delta\Phi_j(-q)} \:,
\end{equation}
where $i$ and $j$ denote the group and (in the case of gluon fields) the
Lorentz
indices. ${\cal R}_k$ stands for the matrix of cutoff terms,
\begin{equation} {\cal R}_k = \left(\begin{array}{*{3}{c}}
   \tilde{R}_k & 0 & 0 \\ 0 & \tilde{R}_k & 0 \\ 0 & 0 & R_k \end{array}\right)
\:,
\end{equation}
written in a block matrix notation. STr denotes the supertrace,
\begin{equation} \mbox{STr}(A) = \mbox{Tr}(M\cdot A)
\end{equation}
for some (super)matrix $A$, where
\begin{equation} M = \left(\begin{array}{*{3}{r}}
   -1 & 0 & 0 \\ 0 & -1 & 0 \\ 0 & 0 & 1 \end{array}\right)
\end{equation}
in the above block matrix notation. Again, the trace includes summation over
the
momenta.

Flow equations for the one--particle irreducible n--point functions can be
derived from (\ref{floweq}) by taking derivatives with respect to the gluon and
ghost fields as well as to the external sources $K$, $L$ and $\bar{L}$,
setting
them equal to zero
after differentiation as usual. The complete set of flow equations is
then represented by an
infinite system of coupled differential equations for the n--point
functions. It
is possible to visualize them in the form of Feynman diagrams. The
$k$--derivative
of a n--point function is given by a finite sum of one--loop graphs, where in
contradistinction to the usual diagrams of perturbation theory the propagators
and vertices are dressed. All the diagrams are infrared and ultraviolet finite
provided the cutoff functions (\ref{cutoff}) appearing in (\ref{floweq})
are appropriately chosen. We
emphasize that despite their one--loop appearance the equations are still
exact.

The introduction of the cutoff term $\Delta S_k$ in the action breaks the BRST
invariance for $\bar{L} = 0$ explicitly. Using the invariance of the functional
integration measure the symmetry breaking can be quantified and expressed in
the
form of modified STI for $\hat{\Gamma}_k$ \cite{E1}:
\begin{equation} \left.\frac{\delta\hat{\Gamma}_k}{\delta K}\cdot
   \frac{\delta\hat{\Gamma}_k}{\delta A} - \frac{\delta\hat{\Gamma}_k}{\delta
L}
   \cdot\frac{\delta\hat{\Gamma}_k}{\delta c} - \frac{\delta\hat{\Gamma}_k}
   {\delta\bar{L}}\cdot\frac{\delta\hat{\Gamma}_k}{\delta\bar{c}}
   \right|_{\bar{L} = 0} =
   \left.\mbox{STr}\left[{\cal R}_k\cdot\frac{\delta^2\hat{\Gamma}_k}{\delta Q
\delta\Phi}\cdot\left(\frac{\delta^2\hat{\Gamma}_k}{\delta\bar{\Phi}\delta\Phi}
   + {\cal R}_k\right)^{-1}\right]\right|_{\bar{L} = 0} \:. \label{STI}
\end{equation}
The external supersources $Q$ have been introduced in (\ref{ssource}). In
(\ref{STI})
the left hand side represents the usual variation under BRST transformations,
while the right hand side specifies the symmetry breaking term, vanishing at $k
= 0$.
Both sides are to be taken at $\bar{L} = 0$.\footnote{A different but
equivalent
formulation is presented in \cite{B,BDM}. The modified STI are called
``fine-tuning conditions'' there.}

The modified STI (\ref{STI}) are similar in form to the flow equations
(\ref{floweq}).
Expanded in the fields and external sources the former are also equivalent to
an infinite
system of coupled nonlinear equations for the $k$--dependent one--particle
irreducible
n--point functions. Again, the trace term can be represented as a finite sum of
one--loop
diagrams with dressed vertices and propagators. Infrared and ultraviolet
finiteness is guaranteed by appropriate choices for the cutoff functions.

Another useful identity is \cite{B}--\cite{E1}
\begin{equation} \partial\cdot\frac{\delta\hat{\Gamma}_k}{\delta K} =
   \frac{\delta\hat{\Gamma}_k}{\delta\bar{c}} \:, \label{DSchw}
\end{equation}
where $\partial$ stands for differentiation with respect to the space--time
variable.
(\ref{DSchw}) is obtained easily by performing a redefinition of the field
$\bar{c}$ in
(\ref{genf}).

The $\bar{L}$--dependence of $\hat{\Gamma}_k$ is trivially fixed to be
\begin{equation} \frac{\delta\hat{\Gamma}_k}{\delta\bar{L}} = \frac{1}{\alpha}
   \partial\cdot A \:. \label{triv}
\end{equation}
The last two identities are already valid at the classical level, i.e.\ with
$\hat{\Gamma}_k$ replaced by $S$. (\ref{DSchw}) and (\ref{triv}) then simply
state the absence of quantum corrections to the classical relations when
expressed
in terms of $\hat{\Gamma}_k$.

The crucial point, as shown in \cite{E1}, is that the flow equations respect
the
modified STI, as well as (\ref{DSchw}) and (\ref{triv}), in the following
sense:
Once the latter are satisfied at a certain scale
$k = \bar{k}$, they are also satisfied for all $k < \bar{k}$ provided
$\hat{\Gamma}_k$ is obtained from $\hat{\Gamma}_{\bar{k}}$ by integration of
the flow
equations. In particular, at $k = 0$ the physical effective action
$\hat{\Gamma}_0$
fulfills the usual STI. Hence the meaning of the modified STI for
$\hat{\Gamma}_k$ is
that they imply the BRST invariance of the physical effective action at $k =
0$.

We remark that the modified STI, as well as the flow equations, can be
extended trivially to include matter fields.

\section{A simple approximation}

The physical quantity one is interested in is the effective action
$\hat{\Gamma}_0$.
Let us briefly outline how the flow equations together with the modified STI
can
be used to calculate $\hat{\Gamma}_0$ in principle. The general idea is to
begin
with a certain $\hat{\Gamma}_{\bar{k}}$ at a scale $\bar{k}$ which has to be
large in comparison with all physical scales of the theory, and then to
integrate
the flow equations down to $k = 0$.
Although the exact form of $\hat{\Gamma}_{\bar{k}}$ is not known a priori, we
expect
from universality
that the resulting $\hat{\Gamma}_0$ is widely independent of the starting point
$\hat{\Gamma}_{\bar{k}}$. In the present case of an asymptotically free theory
the
relevant parameters of $\hat{\Gamma}_{\bar{k}}$ are the coefficients with
non--negative mass dimension in an expansion of $\hat{\Gamma}_{\bar{k}}$ in
fields
and momenta.\footnote{We include here the parameters which are usually termed
marginal.} In addition it is essential for $\hat{\Gamma}_{\bar{k}}$ to
possess the right symmetry properties. In our case of BRST invariance this
is accomplished by imposing the modified STI.

The main problem in practical calculations is of course the integration of the
flow
equations. Since they represent an infinite system of coupled differential
equations
for the n--point functions, an exact
solution seems impossible. Instead one is forced to truncate the system
by taking into account only a finite number of parameters representing the
$k$--dependent effective action $\hat{\Gamma}_k$. In practice one then inserts
the
truncated representation of $\hat{\Gamma}_k$ in the r.h.s.\ of (\ref{floweq})
and
projects out the contributions to the flow of the parameters under
consideration on
the l.h.s. Similarly the modified STI represent an infinite system of coupled
nonlinear
equations for the n--point
functions. By using the truncated form of $\hat{\Gamma}_k$ in (\ref{STI}) and
performing appropriate projections one arrives at a set of equations for the
parameters representing $\hat{\Gamma}_k$ at every scale $k$.

Now taking a functional $\hat{\Gamma}_{\bar{k}}$ which fulfills the modified
STI
exactly and integrating the flow equations down to a lower scale $k$ exactly,
we would obtain some $\hat{\Gamma}_k$ which again satisfies the modified STI.
However, due to the necessary approximations described above this is no longer
true in practice. As a consequence we cannot expect the physical effective
action
$\hat{\Gamma}_0$ to be BRST--invariant if we just integrate the truncated flow
equations.

Let us look at this problem more closely in the special case of the gluon mass.
As will be shown below, for a pure SU(3)--Yang--Mills theory the modified STI
demand a gluon mass term in $\hat{\Gamma}_k$ as long as $k \neq 0$. Stated
differently, for the physical effective action at $k = 0$ to be
BRST--invariant,
the effective action at a scale $k \neq 0$ must contain a precisely specified
gluon mass term. Due to the truncations, however, the gluon mass will deviate
from
the value prescribed by the modified STI in the course of the integration of
the
flow equations, even if one started with the correct value at a higher scale.
Since the mass is a relevant parameter, one will thus not end up with a
BRST--invariant effective action at $k = 0$.

These problems are solved if we impose the truncated form of the
modified STI on $\hat{\Gamma}_k$ at {\em every} scale $k$. In this way we are
`near' to the $\hat{\Gamma}_k$ satisfying the exact modified STI during the
whole
flow and our approximation to the physical effective action
$\hat{\Gamma}_0$ is as close as possible to exact BRST invariance within our
restricted
space of parameters. Since in an exact calculation the modified STI would be
fulfilled
automatically given the correct starting point, we can in fact turn the
argument around
and consider the deviations due to the truncations as providing an internal
validity
check for the approximation.

To be more precise, we have to divide our set of parameters into `dependent'
and
`independent' ones, where the dependent parameters are entirely fixed by the
modified
STI in terms of the independent ones. The number of independent parameters is
the same
as in perturbation theory with BRST--invariant regularization. Now we use the
truncated flow equations to determine the flow of the independent
parameters, but rely on the modified STI to obtain the flow of the dependent
ones.
On the other hand, the flow of the dependent parameters could also be read off
directly from the flow equations without recourse to the modified STI. Due to
the
truncations, the two results obtained in this way will in general differ
slightly. This difference can in turn be regarded as a measure of the validity
of the
approximation.

After this general discussion let us now return to the special case of a
SU(3)--Yang--Mills theory. The classical action including external source terms
was
given in (\ref{class}). A natural ansatz for a truncated $k$--dependent
effective
action would be to include all perturbatively relevant couplings, i.e.\ all
couplings
with non--negative mass dimension.
So we choose as a truncation
\begin{eqnarray}
   \hat{\Gamma}_k & = & \int\mbox{d}^4 x\left\{\frac{Z^A_k}{4}
   F^a_{\mu\nu} F^a_{\mu\nu} + \frac{Z^A_k}{2\alpha_k}\partial_\mu A^a_\mu
   \partial_\nu A^a_\nu + \frac{Z^A_k}{2} m^2_k A^a_\mu A^a_\mu
   \right.\nonumber \\
   & & {}+ Z^c_k\partial_\mu\bar{c}^a\left(\partial_\mu c^a + (Z^A_k)^{1/2}
g^K_k
   f^a{}_{bc} A^b_\mu c^c \right) - Z^c_k K^a_\mu\left(\partial_\mu c^a +
(Z^A_k)^{1/2}
   g^K_k f^a{}_{bc} A^b_\mu c^c\right)  \nonumber \\
   & & \left. {}- \frac{Z^c_k (Z^A_k)^{1/2}}{2} g^L_k L^a
   f^a{}_{bc} c^b c^c + \frac{1}{\alpha^{(0)}} \bar{L}^a\partial_\mu A^a_\mu
\right\}
   \:, \label{quant}
\end{eqnarray}
where
\begin{equation}
   F^a_{\mu\nu} = \partial_\mu A^a_\nu - \partial_\nu A^a_\mu
   + (Z^A_k)^{1/2} g^A_k f^a{}_{bc}A^b_\mu A^c_\nu \:. \label{quantF}
\end{equation}
A priori one would expect the coefficients of all the terms in (\ref{quant}) to
be
independent quantities. However, some of them are related, as will be
explained in detail below. For similar reasons certain terms
are omitted which would be present from dimensional reasoning alone.
Furthermore, powers of the wave function renormalization constants $Z^A_k$ and
$Z^c_k$ for the gluon and ghost fields have been extracted from the coupling
constants, which will prove to be convenient in the course of the calculations.

Let us first consider the coefficient of the $\bar{L} A$--vertex, which
is fixed to be the classical one by (\ref{triv}). For clarity we denote the
`classical' gauge fixing constant $\alpha$ by $\alpha^{(0)}$ in the following.
It appears in the BRST transformation (\ref{BRST})
and through the $\bar{L} A$--vertex enters in the modified STI.
Note that the gauge fixing constant in $\hat{\Gamma}_k$, i.e.\ the coefficient
of the longitudinal
part of the gluon two--point function, will in general be $k$--dependent,
whereas $\alpha^{(0)}$ remains fixed during the flow.

The identity (\ref{DSchw})
demands that $\hat{\Gamma}_k$ depend only on the combination
$\partial\bar{c} - K$, which implies the equality of the corresponding
couplings.
In particular the formally relevant coefficients of the $\bar{c}cAA$-- and the
$\bar{c}c\bar{c}c$--vertex are related to the irrelevant $KcAA$-- and
$KcKc$--couplings respectively and are thereby also rendered irrelevant. Hence
they
are not included in our ansatz (\ref{quant}). The same property of
$\hat{\Gamma}_k$
prevents the appearance of a mass term for the ghost field.\footnote{Observe
however that the cutoff function (\ref{cutoffc}) does introduce a mass--like
term $\sim k^2$ for the ghost field.}

We have introduced a $k$--dependent gauge fixing constant $\alpha_k$ and a
gluon
mass $m_k$. The coefficients of the
$AAA$--, the $KcA$-- and the $Lcc$--vertex are in general different and have
been
denoted by the three coupling constants $g^A_k$, $g^K_k$ and $g^L_k$. We would
like to remark that due to the relations between the different couplings
implied by the modified STI the number of independent couplings is equal to
the number of independent couplings in the relevant part of a BRST--invariant
action. In particular we shall show in the next section how the modified STI
fix the
couplings $m^2_k$, $\alpha_k$, $g^K_k$ and $g^L_k$ in terms of $g^A_k$ and the
product $Z^A_k \alpha^{(0)}$ for every scale $k$.

As implied by (\ref{quantF}), the coupling constant of the four--gluon vertex
is
given by $(g^A_k)^2$. This is a further approximation which is not justified by
the modified STI. In fact these predict a deviation of the four--gluon coupling
constant from the value $(g^A_k)^2$. However we shall see in the next section
that within the region of validity of our approximation the difference
between the coupling constants $g^A_k$, $g^K_k$ and $g^L_k$ may be neglected
for calculational purposes. We therefore expect the same to hold true for the
deviation of the four--gluon coupling constant from $(g^A_k)^2$.

Furthermore the modified STI enforce the appearance of an additional
four--gluon coupling with the group--tensorial structure
\begin{equation}
   \delta_{ab}\delta_{cd} + \delta_{ac}\delta_{bd} + \delta_{ad}\delta_{bc} \:.
\end{equation}
The same structure is also generated by the flow equations, as it of course
must
be for consistency reasons. We neglect this coupling in our approximation
because
it does not contribute directly to the flow of the coupling constants
$g^A_k$, $g^K_k$ and
$g^L_k$, which are our primary objects of interest. In fact it only contributes
to the flow of the gluon mass and the four--gluon coupling itself. Concerning
these additional approximations one should, however, always bear in mind that
the
comparison of the flow of the dependent parameters taken directly from the flow
equations with the one determined via the modified STI provides us with a
measure of the validity of the used truncation, thus indicating every eventual
breakdown of the approximation. This concludes the discussion of the ansatz
for $\hat{\Gamma}_k$ used in the following.

To complete the description of the set--up used for the calculations in the
next
section we need to specify the form of the cutoff functions. These functions
may,
and in our case will, contain the $k$--dependent parameters appearing in
$\hat{\Gamma}_k$. We choose
\begin{eqnarray} R_{k,\mu\nu}(p)
   &=& Z^A_k \left\{\frac{(p^2 + m^2_k)e^{-(p^2 + m^2_k)/k^2}}
   {1 - e^{-(p^2 + m^2_k)/k^2}}\left(\delta_{\mu\nu} - \frac{p_\mu p_\nu}{p^2}
   \right)\right. \nonumber\\
   & & \left.{}+ \frac{(p^2/\alpha_k + m^2_k)e^{-(p^2/\alpha_k + m^2_k)/k^2}}
   {1 - e^{-(p^2/\alpha_k + m^2_k)/k^2}}\frac{p_\mu p_\nu}{p^2}\right\}
   \label{cutoffA}
\end{eqnarray}
as the cutoff function for the gluons and
\begin{equation}
   \tilde{R}_k(p) = Z^c_k \frac{p^2 e^{-p^2/k^2}}{1 - e^{-p^2/k^2}}
   \label{cutoffc}
\end{equation}
for the ghost field. They have the demanded properties, namely to introduce
infrared cutoffs in the respective propagators and to fulfill the conditions
(\ref{cond1}) and (\ref{cond2}), provided that $\alpha_k \ge 0$ and
$\lim_{k \to 0} m^2_k \ge 0$ (typically one finds  $\lim_{k \to 0} m^2_k = 0$,
cf.\ section 4).

The choice (\ref{cutoffA}) may appear unnecessarily complicated at first sight.
However this particular form is chosen to give the propagator appearing in the
loop integrals generated by the trace terms in (\ref{floweq}) and (\ref{STI}) a
fairly simple structure. Employing our ansatz for $\hat{\Gamma}_k$
we obtain for the gluon propagator
\begin{equation}
   \frac{1}{Z^A_k} \left\{\frac{1 - e^{-(p^2 + m^2_k)/k^2}}{p^2 + m^2_k}
   \left(\delta_{\mu\nu} - \frac{p_\mu p_\nu}{p^2}
   \right) + \frac{1 - e^{-(p^2/\alpha_k + m^2_k)/k^2}}{p^2/\alpha_k + m^2_k}
   \frac{p_\mu p_\nu}{p^2}\right\}\delta^{ab} \:. \label{propA}
\end{equation}
A similar calculation for the ghost propagator yields
\begin{equation}
   \frac{1}{Z^c_k} \frac{1 - e^{-p^2/k^2}}{p^2}\delta^{ab} \:. \label{propc}
\end{equation}
Due to the particular structure of the propagators we are able to perform all
loop
integrals appearing in our approximation analytically. However, elegance often
comes
at a price. In our case the drawback is that the cutoff terms involve the
$k$--dependent parameters $Z^A_k$, $Z^c_k$, $m^2_k$ and $\alpha_k$. Hence on
the
r.h.s.\ of the flow equations the $k$--derivatives of all these
parameters appear on evaluating $\partial_k {\cal R}_k$, as will be seen
explicitly in
the next section.

Note that the functions appearing in (\ref{cutoffc}) and (\ref{propc}) as well
as
the coefficients of the transverse and longitudinal momentum projectors in
(\ref{cutoffA}) and (\ref{propA}) are analytical functions of $p^2$, even if
$m^2_k < 0$. In particular they do not contain poles. The limits $p^2 \to 0$
and $m^2_k \to 0$ in (\ref{cutoffA}) are, despite appearance, well--defined and
interchangeable. This guarantees that the functional $\Gamma_k = \hat{\Gamma}_k
+
\Delta S_k$ is local for all values of $m_k$, i.e.\ the n--point functions can
be
expanded in powers of the momenta, which should be the case for a sensible
infrared cutoff.

One might be tempted to choose different mass parameters $m_k$ in the
transverse
and the longitudinal parts in (\ref{quant}) and (\ref{cutoffA}) and hence also
in (\ref{propA}), since it is only the longitudinal part which will be fixed by
the modified STI. However, in this case the coefficient of the non--local term
$p_\mu p_\nu/p^2$
in (\ref{quant}) and (\ref{cutoffA}) would not vanish in the limit $p^2 \to 0$,
and
$\Gamma_k$ would not be local.

\section{Solving the truncated flow equations}

We shall now turn to the integration of the flow equations using the concepts
and
the approximations explained in detail in the previous section. Employing the
ansatz (\ref{quant}) the flow equations and the modified STI
take on a characteristic form which we shall sketch in the following.

We expand the modified STI in the fields and external sources to obtain
equations
for the one--particle irreducible n--point functions. Expanding these in turn
in
powers of momenta and performing appropriate projections we arrive at the
following
set of equations:
\begin{eqnarray}
\mu_k \equiv  \frac{m^2_k}{k^2} &=& \frac{3 g^2_k}{(4\pi)^2} \left(
f^\mu_1(\mu_k,\alpha_k) +
   \frac{\alpha_k}{Z^A_k \alpha^{(0)}} f^\mu_2(\mu_k,\alpha_k) \right)
\nonumber\\
  \frac{\alpha_k}{Z^A_k \alpha^{(0)}} - 1 &=& \frac{3 g^2_k}{(4\pi)^2} \left(
   f^\alpha_1(\mu_k,\alpha_k) + \frac{\alpha_k}{Z^A_k \alpha^{(0)}} f^\alpha_2
   (\mu_k,\alpha_k) \right) \nonumber\\
  \frac{g^A_k - g^K_k}{g^K_k} &=& \frac{3 g^2_k}{(4\pi)^2} \left( f^A_1
   (\mu_k,\alpha_k) + \frac{\alpha_k}{Z^A_k \alpha^{(0)}} f^A_2(\mu_k,\alpha_k)
   \right) \nonumber\\
  \frac{g^L_k - g^K_k}{g^K_k} &=& \frac{3 g^2_k}{(4\pi)^2} \left( f^L_1
   (\mu_k,\alpha_k) + \frac{\alpha_k}{Z^A_k \alpha^{(0)}} f^L_2(\mu_k,\alpha_k)
   \right) \label{STIsyst} \:.
\end{eqnarray}
Here we have introduced the dimensionless mass parameter $\mu_k = m^2_k/k^2$.
The
$f^i_j$ are complicated functions of $\mu_k$ and $\alpha_k$ and do
not depend explicitly on $k$. Hence $k$ enters in the system explicitly only
through
the relation between $m^2_k$ and $\mu_k$.

Furthermore we have made still another approximation by equating the three
coupling
constants $g^A_k$, $g^K_k$ and $g^L_k$ on the r.h.s.\ and
denoting them collectively by $g_k$. Without this approximation we would have
to
replace the r.h.s.\ of (\ref{STIsyst}) by a sum of similar expressions with
$g^2_k$
replaced in turn by the products $(g^A_k)^2$, $g^A_k g^K_k$, $\ldots$ of the
three
coupling constants. The factor of three in front of $g^2_k$ reflects the number
of colors, while $1/(4\pi)^2$ originates from the integration measure of the
loop
integrals.

We see that in the perturbative limit of vanishing coupling constants $g_k \to
0$
the modified STI lead to the usual relations $m^2_k = 0$,
$Z^A_k/\alpha_k = 1/\alpha^{(0)}$ and
$g^A_k = g^K_k = g^L_k$. The last two equations in (\ref{STIsyst}) with the
three
coupling constants set equal to $g_k$ on the r.h.s.\ can thus be considered as
the
first step in an iterational procedure to determine the values of $g^A_k$,
$g^K_k$
and $g^L_k$. What is more, these two equations make it easy to check whether
the approximation of equating the coupling constants is admissible. In fact we
shall find this approximation to be valid in the region of parameter space we
are
interested in, which in turn is limited by the validity of the general
approximation
(\ref{quant}). Since we consider an asymptotically free theory and the starting
scale
$\bar{k}$ is supposed to be large as compared with the physical scales of the
theory,
e.g.\ the confinement scale, we have $g_{\bar{k}} \ll 1$ and the
approximation of equal coupling constants is certainly acceptable for the
starting
point $\hat{\Gamma}_{\bar{k}}$.

In the same way we could have treated the four--gluon coupling constant as
different from $(g^A_k)^2$ in the ansatz (\ref{quant})
and then relate it to $(g^A_k)^2$ through an equation similar to
(\ref{STIsyst}).
However, we expect the difference between these quantities to be negligible
within the region of validity of our approximation.

We observe that $Z^A_k$ enters in (\ref{STIsyst}) only through the combination
$Z^A_k
\alpha^{(0)}$, and $Z^c_k$ does not appear at all. This simplification was in
fact the
reason for extracting powers of the wave function renormalization constants
from the couplings in (\ref{quant}). By counting the parameters
and the equations --- regarding $g^A_k$, $g^K_k$ and $g^L_k$ as different for
the
moment --- we find that the modified STI in the form (\ref{STIsyst}) fix
$\mu_k$, $\alpha_k$, $g^K_k$ and $g^L_k$ in terms of $g^A_k$ and $Z^A_k
\alpha^{(0)}$.
Thus apart from the wave function renormalization constants we end up with
two independent parameters, corresponding to $g$ and $\alpha$ in perturbation
theory with BRST-invariant regularization.

{}From now on we shall always use the approximation of equal coupling
constants,
i.e.\ $g_k = g^A_k = g^K_k = g^L_k$ unless otherwise stated,
so we only need to consider the first two equations in (\ref{STIsyst}). Given
$\mu_k$
and $\alpha_k$ we obtain a system of quadratic equations for $g^2_k$ and
$1/Z^A_k \alpha^{(0)}$, which
can be solved easily. On the other hand we can calculate $\mu_k$ and $\alpha_k$
numerically as functions of $g_k$ and $Z^A_k \alpha^{(0)}$.

Before we come to the discussion of the solutions of (\ref{STIsyst}) let
us consider the limit $\alpha^{(0)} \to 0$, i.e.\ the Landau gauge. One can
show that
in this limit the second equation in (\ref{STIsyst}) implies
$\alpha_k/Z^A_k \alpha^{(0)} = 1$ irrespective of the values of $g_k$ and
$\mu_k$.
Hence $\alpha_k = 0$ and the first equation in (\ref{STIsyst}) becomes
\begin{equation}
   \mu_k = \frac{3 g^2_k}{(4\pi)^2} \left( f^\mu_1(\mu_k,0) + f^\mu_2(\mu_k,0)
   \right) \label{STImu} \:,
\end{equation}
which is much easier to work with. Also the functions $f^i_j$ take on simpler,
though still not enlightening forms in this limit.

We shall now turn to the discussion of the solutions of (\ref{STIsyst}). It is
easy to obtain solutions for small $g_k$ iteratively. Neglecting terms of the
order of $g^4_k$, the first two equations in (\ref{STIsyst}) become
\begin{eqnarray}
  \mu_k &=& \frac{3 g^2_k}{(4\pi)^2} \frac{3(\alpha_k - 1)}{8(\alpha_k + 1)}
  \nonumber \\
  \frac{1}{\alpha_k} - \frac{1}{Z^A_k \alpha^{(0)}} &=& \frac{3
g^2_k}{(4\pi)^2}
  \frac{7 - 10 \alpha_k - 5 \alpha^2_k}{48(\alpha_k + 1)^2} \:. \label{STIonel}
\end{eqnarray}
For brevity we have used the quantity $\alpha_k$ on the r.h.s. However it can
be
replaced by $Z^A_k \alpha^{(0)}$ in the given order in $g_k$ as follows from
the
second equation in (\ref{STIonel}).

It appears problematic
at first sight that the mass parameter $\mu_k$ becomes
negative for $\alpha_k < 1$. However in the limit of vanishing momentum the
gluonic two--point function obtained from $\Gamma_k$ becomes
\begin{equation}
   Z^A_k \frac{\mu_k k^2}{1 - e^{-\mu_k}}\delta_{\mu\nu}
\:,
\end{equation}
which is well--defined and non--negative for all values of $\mu_k$.
Since the propagator
(\ref{propA}) is given by the inverse of this two--point function, the
effective
gluon mass is seen to be non--negative.

We shall refer to (\ref{STIonel}) as the one--loop STI in the following since
they
are valid up to second order in $g_k$. We remark that our expression for
$\mu_k$ differs
from the one--loop result obtained in \cite{E1}, where a different cutoff
function
has been used. Hence the one--loop STI are non--universal in the sense that
they depend on the choice of the cutoff terms.

The solutions of (\ref{STIsyst}) for general values of $g_k$ are shown in
figure 1,
where we have plotted $\mu_k$ as a function of the strong fine--structure
constant
$\alpha_S = g^2_k/4\pi$ for fixed values of $\alpha_k$.
The reason for using $\alpha_k$ as a parameter instead of $Z^A_k \alpha^{(0)}$
is just technical, but since both quantities are $k$--dependent, the use of
$Z^A_k \alpha^{(0)}$ would not be more advantageous anyway. The values of
$\alpha_k$
depicted in figure 1 are, from bottom to top, $\alpha_k = 0$, 1 and 3.

We shall see below that our approximation breaks down at $\alpha_S \approx
1.4$, so
effects showing up at much greater values of $\alpha_S$ are considered as
potential
artifacts of the truncation. We finally remark that $\mu_k$ does not vanish for
$g_k > 0$ in the `Feynman gauge' $\alpha_k = 1$, although it does so
in the one--loop approximation (\ref{STIonel}).

After the discussion of the modified STI let us now turn to the flow equations.
Expanding (\ref{floweq}) in the fields and external sources we obtain flow
equations
for the one--particle irreducible n--point functions. Some of these are
represented
diagrammatically in figure 2, where our ansatz for $\hat{\Gamma}_k$ has already
been
taken into account in that on the r.h.s.\ all vertices not appearing in
(\ref{quant})
are omitted. Since the cutoff functions depend on $k$ not only explicitly, but
also
implicitly through the parameters $Z^c_k$, $Z^A_k$, $m^2_k = \mu_k k^2$ and
$\alpha_k$,
we have
\begin{eqnarray}
  \partial_k R_{k,\mu\nu} &=& \frac{\partial R_{k,\mu\nu}}{\partial k} +
   \frac{\partial R_{k,\mu\nu}}{\partial \ln Z^A_k} \partial_k \ln Z^A_k +
   \frac{\partial R_{k,\mu\nu}}{\partial \mu_k} \partial_k \mu_k +
   \frac{\partial R_{k,\mu\nu}}{\partial \alpha_k} \partial_k \alpha_k
\nonumber \\
  \partial_k \tilde{R}_k &=& \frac{\partial \tilde{R}_k}{\partial k} +
    \frac{\partial \tilde{R}_k}{\partial \ln Z^c_k} \partial_k \ln Z^c_k
\label{dr} \:
\end{eqnarray}
on the r.h.s. of (\ref{floweq}).
 Now we employ our ansatz for $\hat{\Gamma}_k$, expand the flow equations
for the n--point functions in powers of external momenta and perform
appropriate
projections to arrive at equations for the parameters of $\hat{\Gamma}_k$.
Using
the approximation of equal coupling constants $g_k = g^A_k = g^K_k = g^L_k$
discussed
above, the flow equations for the `independent' parameters $Z^A_k$, $Z^c_k$ and
$g_k$
finally take on the following form:
\begin{eqnarray}
  \frac{\partial \ln \lambda_k}{\partial \ln k^2} &=& \frac{3
g^2_k}{(4\pi)^2}\left(
   h^\lambda_0(\mu_k,\alpha_k) + h^\lambda_A(\mu_k,\alpha_k)\frac{\partial \ln
Z^A_k}
   {\partial \ln k^2} + h^\lambda_c(\mu_k,\alpha_k)\frac{\partial \ln Z^c_k}
   {\partial \ln k^2}\right. \nonumber\\
  & & \left.{} + h^\lambda_\mu(\mu_k,\alpha_k)\frac{\partial \mu_k}
   {\partial \ln k^2} + h^\lambda_\alpha(\mu_k,\alpha_k)\frac{\partial
\alpha_k}
   {\partial \ln k^2}\right) \label{floweqsyst} \:,
\end{eqnarray}
where $\lambda_k$ represents the independent parameter. For definiteness we
choose
to take the equation for $\partial \ln g_k/\partial \ln k^2$ from the flow
equation
for the $\bar{c}cA$-- or equivalently the $KcA$--vertex. The functions $h^i_j$
appearing above depend only on the parameters $\mu_k$ and $\alpha_k$ and not
explicitly on $k$.
The terms involving ${\partial \ln Z^A_k}/{\partial \ln k^2}$ etc.
originate from the expressions for $\partial_k R_{k,\mu\nu}$ and
$\partial_k \tilde R_k$ in (\ref{dr}).
Observe that
the classical gauge fixing constant $\alpha^{(0)}$ does not enter in
(\ref{floweqsyst}).

On the r.h.s.\ of (\ref{floweqsyst}) the $k$--derivatives of $\mu_k$ and
$\alpha_k$
appear. Since these parameters are given in terms of $g_k$ and
$Z^A_k \alpha^{(0)}$ through the modified STI, one can obtain expressions for
$\partial \mu_k/\partial \ln k^2$ and $\partial \alpha_k/\partial \ln k^2$ by
taking the $k$--derivative of the first two equations in (\ref{STIsyst}). We
write
them in the form
\begin{eqnarray}
  \frac{\partial \mu_k}{\partial \ln k^2} &=& 2\mu_k\frac{\partial \ln g_k}
   {\partial \ln k^2} + \frac{3 g^2_k}{(4\pi)^2}\left(
h^\mu_A\big(\mu_k,\alpha_k,Z^A_k
   \alpha^{(0)}\big)\frac{\partial \ln Z^A_k}{\partial \ln k^2} \right.
\nonumber \\
  & & \left. {} + h^\mu_\mu\big(\mu_k,\alpha_k,Z^A_k\alpha^{(0)}\big)
   \frac{\partial \mu_k}{\partial \ln k^2} +
h^\mu_\alpha\big(\mu_k,\alpha_k,Z^A_k
   \alpha^{(0)}\big)\frac{\partial \alpha_k}{\partial \ln k^2} \right) \:,
\nonumber \\
  \frac{\partial \alpha_k}{\partial \ln k^2} &=& \alpha_k \frac{\partial \ln
Z^A_k}
   {\partial \ln k^2} + 2\left(\alpha_k -
Z^A_k\alpha^{(0)}\right)\frac{\partial \ln g_k}
   {\partial \ln k^2} + \frac{3 g^2_k}{(4\pi)^2}\left(
h^\alpha_A\big(\mu_k,\alpha_k,
   Z^A_k\alpha^{(0)}\big)\frac{\partial \ln Z^A_k}{\partial \ln k^2} \right.
   \nonumber \\
  & & \left. {} + h^\alpha_\mu\big(\mu_k,\alpha_k,Z^A_k\alpha^{(0)}\big)
   \frac{\partial \mu_k}{\partial \ln k^2} +
h^\alpha_\alpha\big(\mu_k,\alpha_k,
   Z^A_k\alpha^{(0)}\big)\frac{\partial \alpha_k}{\partial \ln k^2} \right) \:.
   \label{flowSTI}
\end{eqnarray}
The functions $h^i_j$, now depending additionally on $Z^A_k\alpha^{(0)}$, are
given by
the functions $f^i_j$ appearing in (\ref{STIsyst}), their derivatives with
respect to
$\mu_k$ and $\alpha_k$, and some factors of $\mu_k$, $\alpha_k$ and
$Z^A_k\alpha^{(0)}$.

(\ref{floweqsyst}) and (\ref{flowSTI}) together represent a system of five
linear
equations for the $k$--derivatives of the five parameters $Z^A_k$, $Z^c_k$,
$g_k$,
$\mu_k$ and $\alpha_k$, where the coefficients are functions of $g_k$, $\mu_k$,
$\alpha_k$ and $Z^A_k \alpha^{(0)}$. Since only two of the latter
parameters are
independent by virtue of the
modified STI, we conclude that the $k$--derivatives of all parameters are fixed
by two independent quantities, say $g_k$ and $Z^A_k \alpha^{(0)}$.

Choosing initial values for
$g_{\bar{k}}$, $\alpha^{(0)}$ (and
$Z^A_{\bar{k}}=1$, $Z^c_{\bar{k}}=1$),
and determining the `dependent' parameters $\mu_{\bar{k}}$ and
$\alpha_{\bar{k}}$ with the help of the modified STI, we can now integrate the
flow
equations using (\ref{floweqsyst}) and (\ref{flowSTI}). We remark that the
division
into dependent and independent parameters is not unique and that we could as
well
regard $\mu_k$ and $\alpha_k$ as independent parameters determining the values
of
$g_k$ and $Z^A_k \alpha^{(0)}$.

Before we present the results of the integration of the flow equations let us,
however,
have a look at the way we can control the consistency of our approximation. In
(\ref{flowSTI}) we have taken
$\partial \mu_k/\partial \ln k^2$ and $\partial \alpha_k/\partial \ln k^2$
from the $k$--derivative of the modified STI,
but alternatively we could use
flow equations derived from (\ref{floweq}) for $\partial \mu_k/\partial \ln
k^2$ and $\partial \alpha_k/\partial \ln k^2$
(as in the case of the other parameters).
Due to the used truncations the corresponding results
will generally be different, and this difference can be
taken as a measure of the validity of the approximation.

Concerning the additional approximation of equal coupling constants, we could
have
taken the equation for $\partial \ln g_k/\partial \ln k^2$ in
(\ref{floweqsyst}) just
as well from the flow equation for the three--gluon or the $Lcc$--vertex. If
the
approximation $g^A_k = g^K_k = g^L_k$ is to hold, we should obtain at least
approximately the same result for $\partial \ln g_k/\partial \ln k^2$
irrespective
of the equation we use. This represents an additional condition to
be fulfilled within the region of validity of our approximation.

Let us briefly discuss the Landau gauge $\alpha^{(0)} = 0$. In this case
$\alpha_k$ is fixed to be zero for all values of $k$ by the modified STI, as
was
remarked above. Hence the modified STI imply that $\alpha_k = 0$ is a
fixed point in the Landau gauge. This is consistent with the flow equations,
which
demand that the $k$--derivative of $\alpha_k$ vanish at $\alpha_k = 0$.

We now turn to the solutions of the flow equations and start with the
discussion
of the one--loop approximation, i.e.\ we neglect terms of the order of $g_k^4$.
Again, it is easy to solve (\ref{floweqsyst}) and (\ref{flowSTI}) iteratively
in the
limit of small $g_k$, and we find
\begin{eqnarray}
  \frac{\partial \ln Z^A_k}{\partial \ln k^2} &=& \frac{3 g^2_k}{(4\pi)^2}
   \frac{13 - 3\alpha_k}{6} \nonumber \\
  \frac{\partial \ln Z^c_k}{\partial \ln k^2} &=& \frac{3 g^2_k}{(4\pi)^2}
   \frac{3 - \alpha_k}{4} \nonumber \\
  \frac{\partial \ln g_k}{\partial \ln k^2} &=& -\frac{3 g^2_k}{(4\pi)^2}
   \frac{11}{6} \nonumber \\
  \frac{\partial \mu_k}{\partial \ln k^2} &=& 0 \nonumber \\
  \frac{\partial \alpha_k}{\partial \ln k^2} &=& \alpha_k \frac{\partial \ln
Z^A_k}
   {\partial \ln k^2} \:, \label{floweqonel}
\end{eqnarray}
where we have made use of the one--loop STI (\ref{STIonel}) in order to
replace $\mu_k$ and $Z_k^A \alpha^{(0)}$ on
the r.h.s. The first three equations
represent the results familiar from perturbation theory, while the last two
state that
the modification of the STI does not affect the $\beta$--functions at the
one--loop
level (remember that $\alpha_k = Z^A_k \alpha^{(0)}$ in perturbation theory
with
BRST--invariant regularization).

Concerning the consistency of our approximation, we find that
the flow equations for $\partial \mu_k/\partial \ln k^2$ and
$\partial \alpha_k/\partial \ln k^2$ derived from (\ref{floweq})
reproduce, to second order in $g_k$, exactly the results in
(\ref{floweqonel}). Here we have
used (\ref{STIonel}) in order to rewrite the r.h.sides.
Hence the flow equations are entirely consistent with
the modified STI at the one--loop level. Likewise the expressions for $\partial
\ln g_k
/\partial \ln k^2$ taken from the flow equations for the three--gluon and the
$Lcc$--vertex are identical with the one in (\ref{floweqonel}) in second order
in $g_k$.
This is of course necessary for our approximation to be valid
within the range of parameters where one expects the one--loop approximation to
be
reliable.

Let us now go beyond the one--loop level. Before we present the full solution,
however, we would like to discuss briefly a different approximation that makes
an
analytical solution of the flow equations possible. It consists in arbitrarily
setting $\mu_k$ equal to zero for all $k$, thereby ignoring the modified STI.
Comparing with the full solution below where the modified STI are taken into
account, we
can thus estimate the effect of neglecting the gluon mass parameter.
Furthermore we work
in the Landau gauge.

Setting $\mu_k$ and $\alpha_k$ to zero, (\ref{floweqsyst}) becomes
\begin{eqnarray}
  \frac{\partial \ln Z^A_k}{\partial \ln k^2} &=& \frac{3
g^2_k}{(4\pi)^2}\left(
   \frac{13}{6} + \frac{1 + 150\ln(4/3)}{36}\cdot
   \frac{\partial \ln Z^A_k}{\partial \ln k^2} - \frac{1 -
36\ln(4/3)}{216}\cdot
   \frac{\partial \ln Z^c_k}{\partial \ln k^2}\right) \nonumber \\
  \frac{\partial \ln Z^c_k}{\partial \ln k^2} &=& \frac{3
g^2_k}{(4\pi)^2}\left(
   \frac{3}{4} + \frac{3\ln(4/3)}{4}\cdot\frac{\partial \ln Z^A_k}{\partial \ln
k^2}
   + \frac{3\ln(4/3)}{4}\cdot\frac{\partial \ln Z^c_k}{\partial \ln k^2}\right)
   \nonumber \\
  \frac{\partial \ln g_k}{\partial \ln k^2} &=& \frac{3 g^2_k}{(4\pi)^2}\left(
   -\frac{11}{6} - \frac{1 + 204\ln(4/3)}{72}\cdot
   \frac{\partial \ln Z^A_k}{\partial \ln k^2} + \frac{1 -
360\ln(4/3)}{432}\cdot
   \frac{\partial \ln Z^c_k}{\partial \ln k^2}\right)
   \label{floweqmuzero} \:.
\end{eqnarray}
In second order in $g_k$ we recover the one--loop results (\ref{floweqonel})
with
$\alpha_k = 0$.

{}From the solution of the system of linear equations (\ref{floweqmuzero}) we
obtain
\begin{equation}
   \frac{\partial \ln g_k}{\partial \ln k^2} = - \frac{2.6 \cdot 3
g^2_k/(4\pi)^2
   \cdot (2.7 - 3 g^2_k/(4\pi)^2)}{(4.8 - 3 g^2_k/(4\pi)^2)(0.81 - 3
g^2_k/(4\pi)^2)} \:,
   \label{solmuzero}
\end{equation}
where we have given approximate numerical values for the appearing
numbers.

Since SU(3)--Yang--Mills theory is asymptotically free, we start with a small
coupling
constant $g_{\bar{k}}$ at some large scale $\bar{k}$. Lowering the scale the
coupling constant increases according to the solution of (\ref{solmuzero})
until it reaches the point $3 g^2_k/(4\pi)^2 = 0.81$, where
the $\beta$--function given by (\ref{solmuzero}) diverges. At this point $k^2$
as a function of $g^2_k$ reaches a maximum, which means that we cannot
integrate the
equation beyond this point. In contrast to perturbation theory, where we run
into
a Landau pole, the solution here just ends at a finite value of $g^2_k$.

In figure 3 we have plotted $\alpha_S = g^2_k/4\pi$ as a function of
$\ln (\bar{k}^2/k^2)$ in this approximation (dot--dashed curve), together with
the
three--loop perturbative result (dotted curve)
and the full solution to be discussed in the following.

Taking the modified STI into account, we determine the flow in the
two--parameter
space $((\alpha_k - 1)/(\alpha_k + 1),g^2_k/4\pi)$ (which maps
 values of $\alpha_k$ between zero and infinity into a finite interval).
 As discussed above, the
$k$--derivatives of all parameters are already fixed by these two independent
ones.
One could have chosen another set as well, e.g.\ $Z^A_k \alpha^{(0)}$ instead
of
the above function of $\alpha_k$, but the present choice turned out to be
particularly
convenient. The flow towards lower values of $k$ in this parameter space
induced
by the flow equations is indicated in figure 4 by a vector field, where each
vector
represents the negative of the derivative
with respect to $\ln k^2$. To make the directions and the magnitudes of all the
vectors in the diagram visible, we had to rescale the vectors in a
slightly unusual way, where
the actual lengths $L$ are replaced by $0.06 + 0.015 \cdot L$.

Starting with small values of $g_k$ because of asymptotic freedom, we see from
figure
4 that the flow is directed towards larger values of $g_k$, as expected from
perturbation theory. The flow in the horizontal direction
depends on the initial value of $\alpha_k$. For
$\alpha_k < 13/3$ the system is driven towards smaller values of $\alpha_k$,
for
$\alpha_k > 13/3$ towards larger ones. The specific value 13/3 can be read off
from the one--loop formula (\ref{floweqonel}). As shown in figure 4, this
behaviour
remains unchanged for larger values of $g_k$.

The curves appearing in the figure represent the boundary of the region
where our approximation is valid. The different curves are obtained by imposing
different constraints on the deviations resulting from our approximation as
discussed above. The approximation can be considered valid below both curves
and
unreliable above. More details on the constraints determining the curves will
be
given below.

We find that for initial values $\alpha_k > 13/3$ the system flows out
of the region
of validity of the approximation quickly, the values of $g_k$ still
being small.
Therefore we concentrate in the following on initial values $\alpha_k <
13/3$. Here
the flow converges asymptotically towards $\alpha_k = 0$, the Landau
gauge case.
So $\alpha_k = 0$ is not only a fixed point of the flow, but even an
attractive one.
This is a very important feature because it shows how independence of
the gauge
parameter emerges in our formalism: Starting with an arbitrary value
$\alpha_{\bar{k}} < 13/3$, corresponding to a certain value of
$Z^A_{\bar{k}} \alpha^{(0)}$, the system flows towards $\alpha_k = 0$,
and hence at a lower scale $k$ the functions $f^i_j$ and $h^i_j$
appearing on the
r.h.s.\ of the modified STI
and the flow equations approach the form they have in the limit
$\alpha_k \to 0$. Thus in the limit $k \to 0$ the parameters in the
effective action become
independent of $\alpha^{(0)}$, provided this feature persists beyond our
reliability bounds.

Let us now discuss the curves in figure 4 representing the boundary of
the region of validity of the approximation. From the above discussion
it is clear
that this region is determined by several conditions. First of all the
deviations
of the coupling constants $g^A_k$ and $g^L_k$ from $g^K_k$ must be
small. We can
estimate the relative deviations from the last two equations in
(\ref{STIsyst}).
Furthermore the $k$--derivatives of the parameters must not change
drastically
when we calculate them in one of the alternative ways described above.
We concentrate
here on the $\beta$--function of the coupling constant, i.e.\ on
$\partial \ln g_k/
\partial \ln k^2$. We require its relative deviation to be small when we
use the
flow equation for the three-gluon or the $Lcc$--vertex instead of the
one for the
$\bar{c}cA$--vertex in (\ref{floweqsyst}), and likewise when we take
$\partial \mu_k
/\partial \ln k^2$ or $\partial \alpha_k/\partial \ln k^2$ from
(\ref{floweq})
 instead of using the $k$--derivative of the modified STI in
(\ref{flowSTI}). Now the solid curve in figure 4 represents the boundary of the
region where all the relative deviations mentioned above are smaller than 3\%,
while
for the region bounded by the dot--dashed curve they are smaller than 10\%.
Both curves
consist of several smooth parts, each of which represents one specific
constraint.

The region bounded by the dot--dashed curve extends to $g^2_k/4\pi \approx 2.8$
at
$(\alpha_k - 1)/(\alpha_k + 1) \approx 0.2$. However to access this region
starting
with a small value $g_{\bar{k}}$ one would have to fine--tune the initial value
$\alpha_{\bar{k}}$. The most interesting region due to the attractivity of the
fixed point $\alpha_k = 0$ is the one with small values of $\alpha_k$. Here the
strongest constraints are obtained from replacing the equation for
$\partial \ln g_k
/\partial \ln k^2$ in (\ref{floweqsyst}) by the one taken from the flow
equation
for the three-gluon vertex for the solid curve and from replacing the
equation for
$\partial \mu_k/\partial \ln k^2$ in (\ref{flowSTI}) by the one taken
from (\ref{floweq})
for the dot--dashed curve. Hence the point where our
approximation is expected to become inappropriate is determined
by the value of $g_k$ where the flow equations and the $k$--derivative of the
modified STI cease to be consistent. Figure 5 shows the $\beta$--function of
$g_k$
at $\alpha_k = 0$ as calculated from (\ref{floweqsyst}) and (\ref{flowSTI})
(dot--dashed
curve) and by taking $\partial \mu_k/\partial \ln k^2$ from
(\ref{floweq})
instead of using the $k$--derivative of the modified STI (solid curve).
More precisely, $\partial \ln \alpha_S/\partial \ln k^2$ is
plotted as a function of $\alpha_S$, where $\alpha_S = g^2_k/4\pi$.

The result of the integrated flow equations is shown in figure 3 in the
case of
the Landau gauge $\alpha_k = 0$. Since $\alpha_k = 0$ is an attractive
fixed point,
the curves look the same for all initial values $\alpha_{\bar{k}}$,
provided the
starting scale $\bar{k}$ is large enough. In figure 3 we have plotted
$g^2_k/4\pi$
as a function of $\ln (\bar{k}^2/k^2)$, so the flow is directed towards
the right.
The initial value for the coupling constant is $g_{\bar{k}} = 1$. The figure
shows, from left to right, the perturbative three--loop result
\cite{tara}, the full
solution
from the flow equations, and the analytical solution in the approximation
$\mu_k = 0$. We see that the three curves are nearly identical as long as
$\alpha_S = g^2_k/4\pi$ is smaller than 0.2 approximately, but already for
$\alpha_S \approx 0.5$, where our approximation
is still expected to be reliable, deviations become visible. We conclude that
the effect of the non--vanishing gluon mass parameter becomes important here.
Observe that the value of $k^2$ where our approximation ceases to be reliable,
i.e.\ where $\alpha_S \approx 1.4$, lies beyond the Landau pole of three--loop
perturbation theory.

Finally, let us have a look at the behaviour of the gluon mass
parameter, again
for $\alpha_k = 0$.
Figure 6 shows the normalized mass parameter $m^2_k/\bar{k}^2$ as a function
of $\ln (\bar{k}^2/k^2)$. It is driven to zero during the flow, and we
emphasize
again that this is nontrivial, since a slightly different starting
point $m^2_{\bar{k}}$ is not expected to lead to a similar behaviour.

\section{Conclusions and Outlook}

To summarize, we have shown that the method proposed in \cite{E1} to use
flow equations
for the description of gauge theories is well suited for practical
calculations. This
has been demonstrated here for the case of a SU(3)--Yang--Mills theory
in a simple but
non--trivial approximation, where the $k$--dependent effective action is
approximated by its perturbatively relevant part.

We have emphasized the crucial role played by the modified STI in this
approach both for
the general concept and for practical calculations, where approximations
are inevitable.
These identities allow to
control the used approximation by providing internal consistency checks.
We believe that this is a new and very useful property, which of course
rests upon the
local gauge symmetry of the original theory.

We have reproduced the usual perturbative one--loop results in the
limit of small coupling constants, and we have studied in detail the
behaviour and the effects of the gluon mass term (in particular on the
higher order results), which is required by the modified STI.
Finally we have argued how independence of
the gauge fixing constant arises in our formalism.

The fact that the internal consistency checks did not allow us to go
much beyond standard perturbation theory is an obvious concequence
of our truncation of the effective action, not of the method per se:
The discrepancies between flow equations and modified STI for small $k$
tell us that higher dimensional operators are required in order to obtain
a consistent description in this regime; this is not astonishing in
view of the expected dynamical effects such as glueballs and condensates.
Higher dimensional operators should allow, e.g., to describe an arbitrary
momentum dependence of propagators and vertices.
In the case of non-gauge theories, such more general parametrizations of
effective actions have already been used for studies of phenomena like
the formation of bound states, condensates, and other non--perturbative
effects \cite{E2,W1}. The obvious next steps are now to
investigate these phenomena within non-abelian gauge theories, using
generalisations of the method developed in the present paper.

\section*{Acknowledgements}

Two of us (M.\ H.\ and A.\ W.) would like to thank the Laboratoire de Physique
Th\'eorique et Hautes Energies for the
kind hospitality and M.\ Reuter and C.\ Wetterich for discussions.

\newpage

\section*{Figure captions}

\newcounter{fig}
\begin{list}{\bf Figure \arabic{fig}:}{\usecounter{fig}}

\item The dimensionless mass parameter $\mu_k = m^2_k/k^2$ as a function of the
strong
  fine--structure constant $\alpha_S = g^2_k/4\pi$ for different values of the
  $k$--dependent gauge fixing constant
  $\alpha_k$. The values depicted in the figure are $\alpha_k = 0$ (solid
curve), 1
  (dot--dashed curve) and 3 (dotted curve).

\item Flow equations for the one--particle irreducible n--point functions. The
vertex
  functions are to be taken from $\hat{\Gamma}_k$, while the propagators are
given in
  (\ref{propA}) and (\ref{propc}). The figure shows the flow equations for the
gluon
  two--point function, the ghost two--point function and the
$\bar{c}cA$--vertex.

\item The strong fine--structure constant $\alpha_S = g^2_k/4\pi$ as a function
of
  the scale $k$ in the Landau gauge $\alpha_k = 0$. The initial value at $k =
\bar{k}$
  is taken to be $g_{\bar{k}} = 1$. The figure depicts the perturbative
three-loop
  result (dotted curve), the full solution from the flow equations (solid
curve) and
  the solution in the approximation $\mu_k = 0$ (dot--dashed curve).

\item Representation of the flow in the two--parameter space $((\alpha_k -
1)/(\alpha_k
  + 1),\alpha_S = g^2_k/4\pi)$. The arrows are given by the negative of the
derivatives
  with respect to $\ln k^2$, their lengths $L$ rescaled to $0.06 + 0.015 \cdot
L$ in the
  figure. The curves represent the boundary of the region of validity of the
  approximation. The solid curve results from the 3\%--constraints and the
  dot--dashed curve from the 10\%--constraints (see section 4 for details).

\item The $\beta$--function $\beta(\alpha_S) = \partial \ln \alpha_S/\partial
\ln k^2$
  as a function of the strong fine--struc\-ture constant $\alpha_S =
g^2_k/4\pi$ in the
  Landau gauge $\alpha_k = 0$. The dot--dashed curve results from
(\ref{floweqsyst}) and
  (\ref{flowSTI}), while for the solid curve $\partial \mu_k/\partial \ln k^2$
is taken
  from (\ref{floweq}). The deviation can be used to determine the region of
validity of
  the approximation.

\item The normalized mass parameter $m^2_k/\bar{k}^2$ as a function of the
scale $k$
  in the Landau gauge $\alpha_k = 0$. As in figure 3, the initial value at $k =
\bar{k}$
  is determined by $g_{\bar{k}} = 1$.

\end{list}

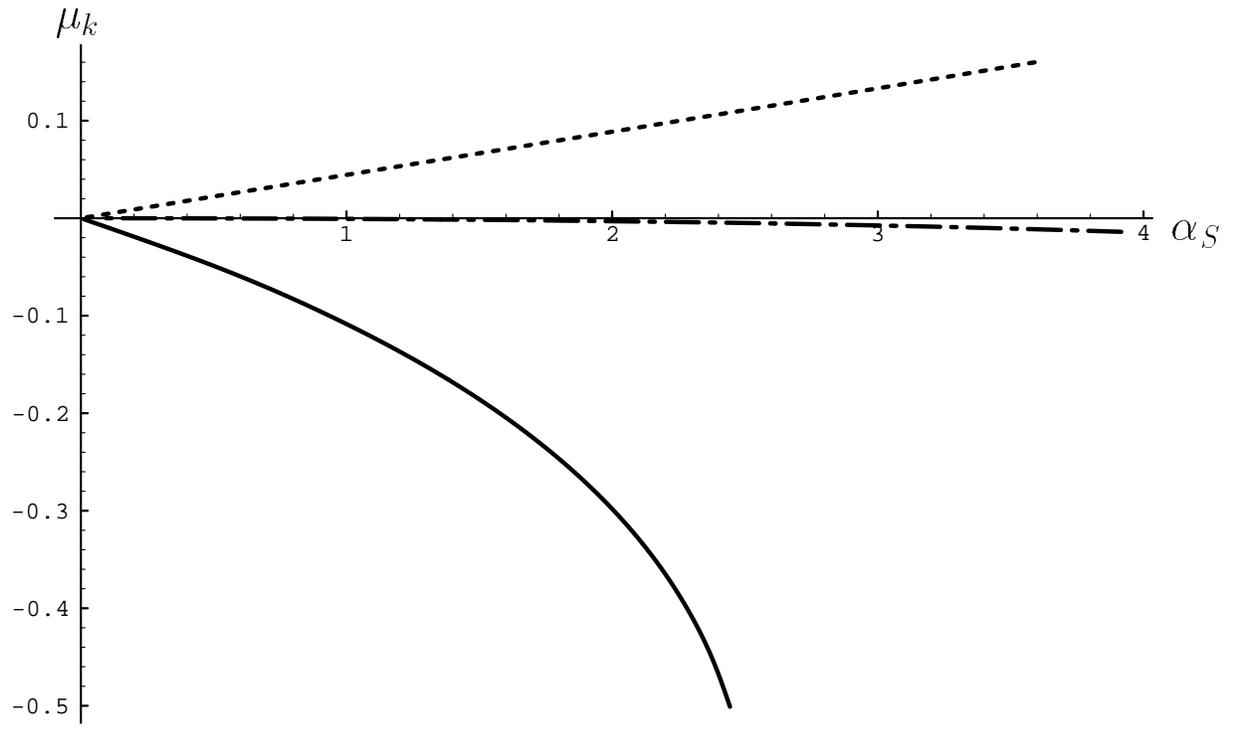
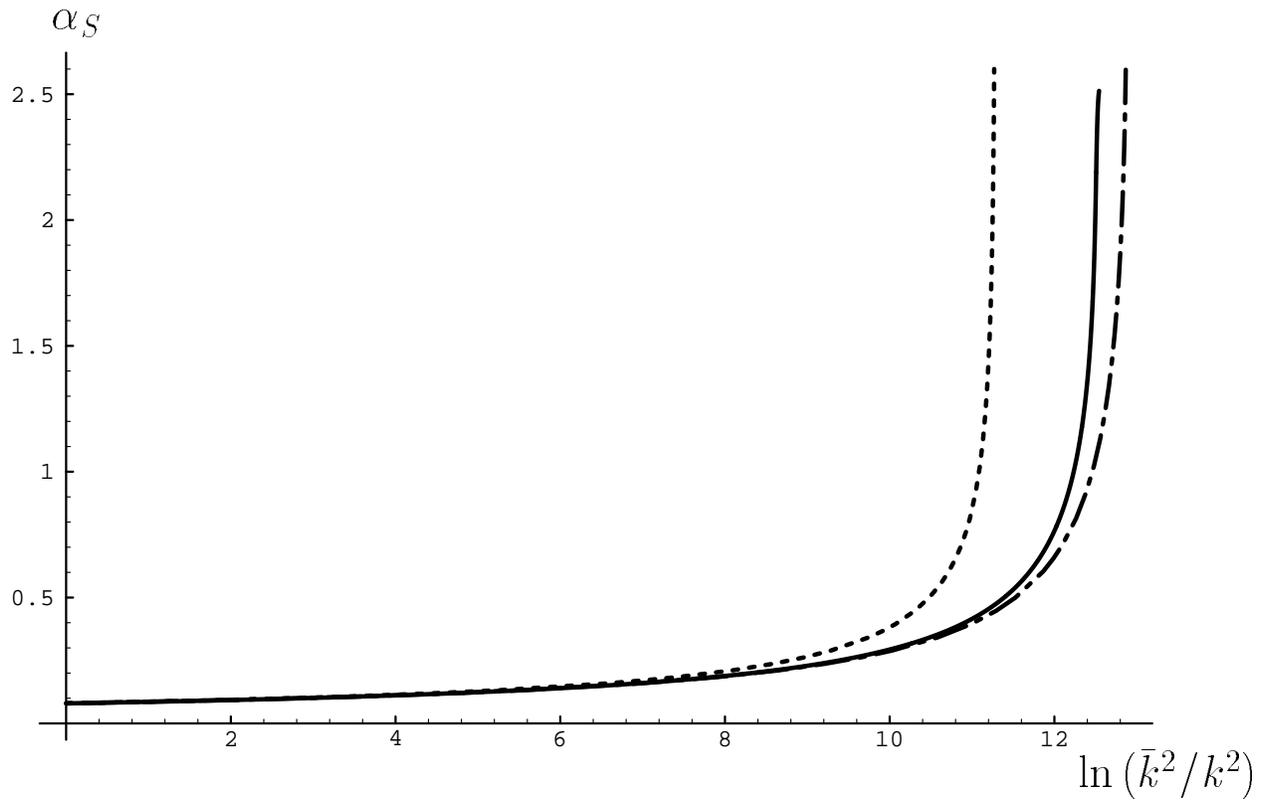
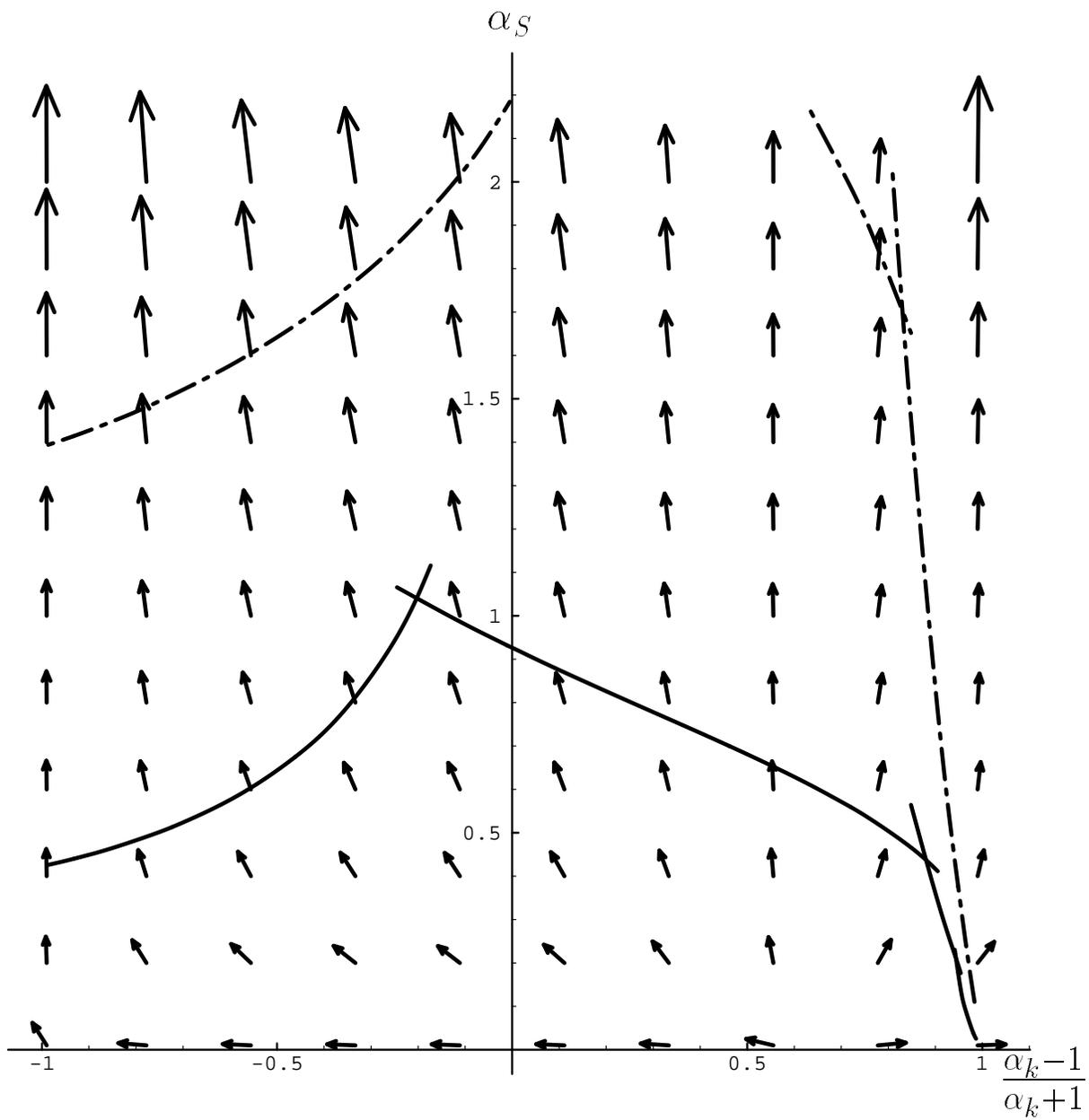
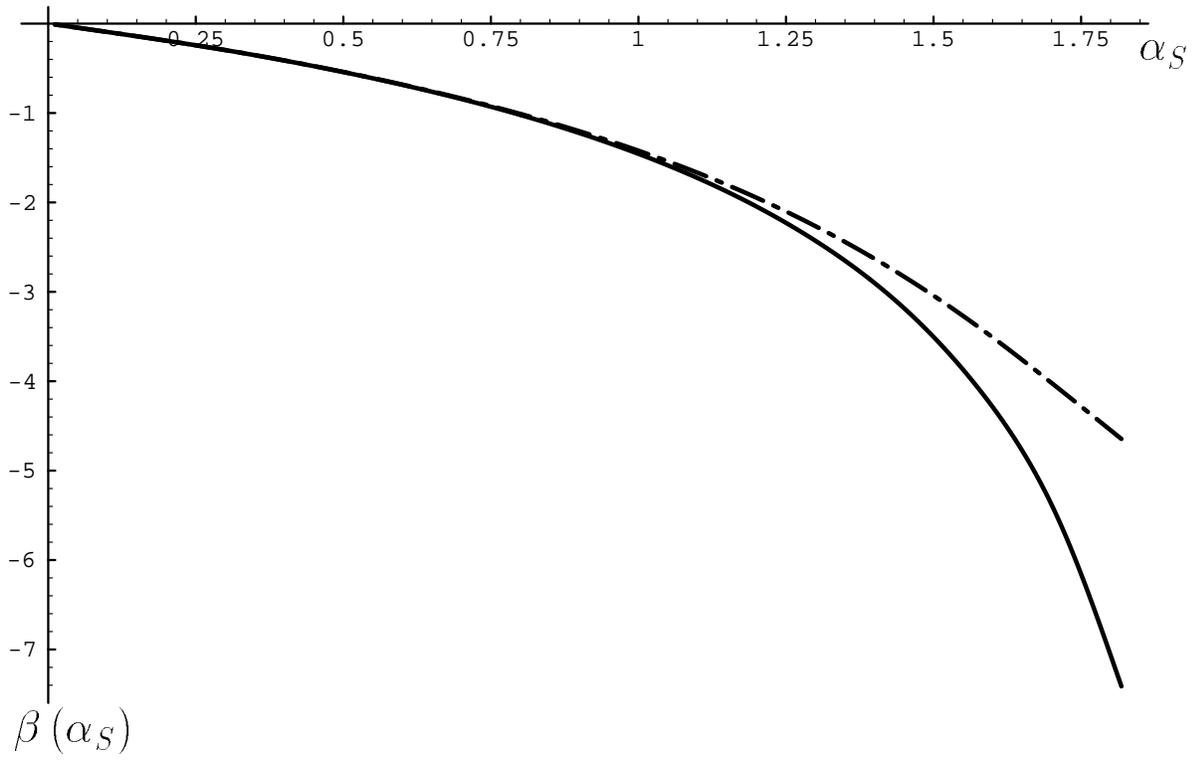
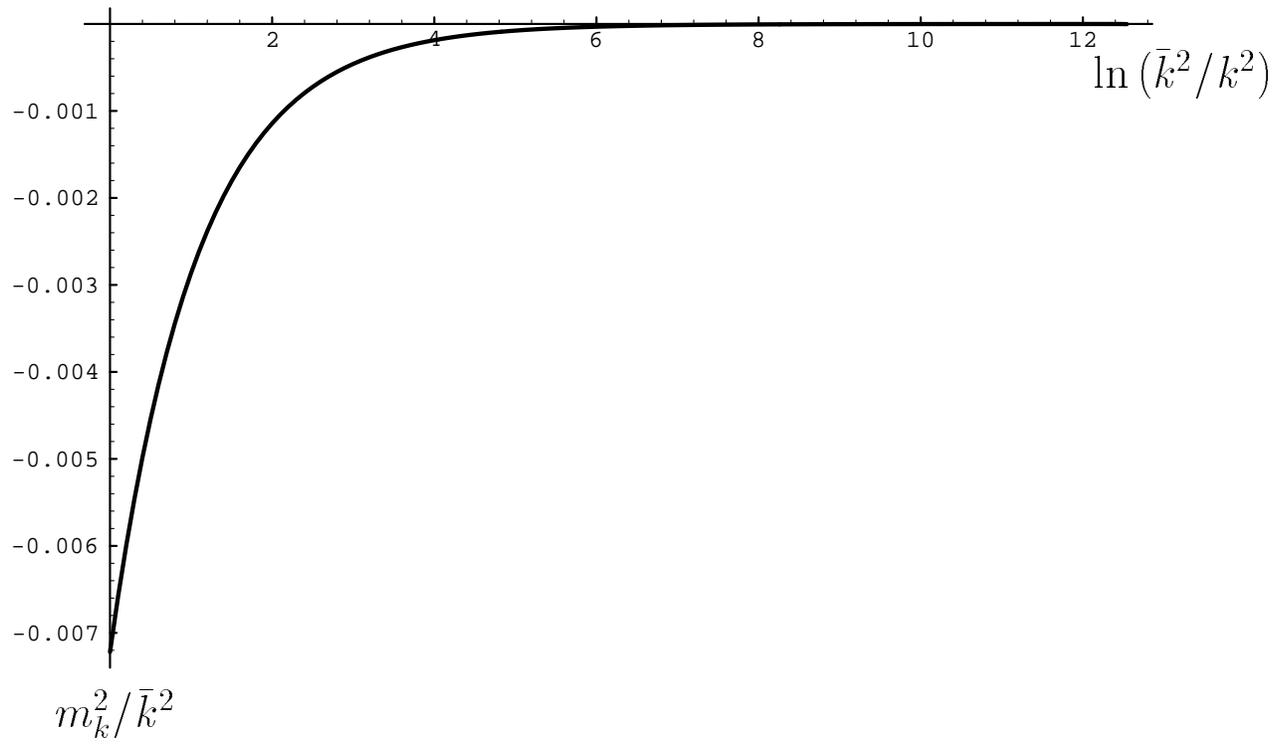
\end{document}